# Anisotropic electrical resistance in mesoscopic LaAlO$_3$/SrTiO$_3$ devices with individual domain walls

Nicholas J. Goble[1], Richard Akrobetu[2], Hicham Zaid[3], Sukrit Sucharitakul[1], Marie-Hélène Berger[3], Alp Sehirlioglu[2], and Xuan P. A. Gao[1,*]

[1]Department of Physics, Case Western Reserve University, Cleveland, Ohio 44106, USA
[2]Department of Materials Science and Engineering, Case Western Reserve University, Cleveland, Ohio 44106, USA
[3]MINES Paris Tech, PSL Research University, MAT - Centre des matériaux, CNRS UMR 7633, BP 87 91003 Evry, France
[*] Email: xuan.gao@case.edu

**ABSTRACT**

The crystal structure of bulk SrTiO$_3$(STO) transitions from cubic to tetragonal at around 105K. Recent local scanning probe measurements of LaAlO$_3$/SrTiO$_3$ (LAO/STO) interfaces indicated the existence of spatially inhomogeneous electrical current paths and electrostatic potential associated with the structural domain formation in the tetragonal phase of STO. However, how these effects impact the electron conduction in LAO/STO devices has not been fully studied. Here we report a study of temperature dependent electronic transport in combination with the polarized light microscopy of structural domains in mesoscopic scale LAO/STO devices. By reducing the spatial size of the conductive interface to be comparable to the size of a single tetragonal domain of STO, the anisotropy of interfacial electron conduction in relationship to the domain wall and its direction was characterized in the temperature range $T$=10-300K. It was found that the four-point resistance measured with current parallel to the domain wall in device is larger than the resistance measured perpendicular to the domain wall. This observation is qualitatively consistent with the current diverting effect from a more conductive domain wall within the sample. Among all the samples studied, the maximum resistance ratio found is at least 10 and could be as large as $10^5$ at $T$=10K. This electronic anisotropy may have implications on other oxide hetero-interfaces and the further understanding of electronic/magnetic phenomena found in LAO/STO.

**Introduction**

The structures and structural phase transitions of SrTiO$_3$ had been a subject of research interests since decades ago[1,2]. Since the discovery of quasi-two-dimensional(q2D) electron conduction between LaAlO$_3$ and TiO$_2$ terminated SrTiO$_3$ in 2004[3], LAO/STO heterointerfaces have quickly become one of the fastest growing research fields in physics and material science research[4-7]. While LAO/STO is being explored for novel nanoelectronic and memory device applications[8-11], researchers also uncovered many exotic phenomena in LAO/STO such as gate tunable conductivity and superconductivity[12-14], magnetism[15] and even the coexistence of magnetism with superconductivity[16-19]. Thanks to the success of LAO/STO, oxide based heterostructures are experiencing a significant boom in research focus[20]. Al$_2$O$_3$/SrTiO$_3$[21,22], LaCoO$_3$/SrTiO$_3$ [23], GdTiO$_3$/SrTiO$_3$[24], and SmTiO$_3$/SrTiO$_3$[25] are just a few of the many exciting materials that have been realized due to this concentration in oxide heterointerfaces.

Along with numerous physical phenomena, perovskite oxide heterointerface systems boast complexities that rival high-$T_c$ superconductors. Oxygen growth pressure[26-28], La/Al stoichometry[29], and cation intermixing[30,31] are a few of known factors that can affect interfacial conduction in LAO/STO. Recently, scanning probe or scanning electron microscopy techniques showed that crystal domain boundaries in the STO layer of LAO/STO modify the local conductivity of the interface and create regions of striped potential modulations [32-34]. However, it is unclear if such microscopic effects would impact the electrical devices based on LAO/STO in any gross way. In an electrical transport study of macroscopic LAO/STO samples with size about 5mm×5mm, Schoofs *et al.* [35] discovered hysteresis effects between the warm up and cool down temperature ($T$) dependent resistance $R(T)$ at $T\sim85K$ as well as $\sim180K$ in samples with q2D electron density a few times $10^{13}/cm^2$ and the q2D confinement is strong. These hysteresis effects were attributed to structural phase transitions whose transition temperature may be different from the bulk due to the surface nature of the LAO/STO. But the exact nature of these two structural transitions were not known. In the present work, we combined the polarized light microscopy imaging of structural domains and direct electrical transport measurement of the same device for the first time and obtained two new findings. First, we identified the hysteretic $R(T)$ ~85K in LAO/STO to be related to the cubic to tetragonal transition as confirmed by in situ polarized light microscopy. The higher temperature hysteresis/transition at ~180K does not induce any feature detectable by polarized microscopy. Moreover, we found that mesoscopic LAO/STO devices containing single tetragonal domain walls exhibit anisotropic electron transport at low temperatures: the resistance ratio between the current parallel to the domain wall versus current perpendicular to the domain wall configurations is at least ten at 10K and could reach as high as $10^5$ in a sample which showed diverging insulating $R(T)$ at low $T$ along the high resistance configuration. In addition to the potential relevance to the understanding of other non-uniform properties of oxide hetero-interfaces, this striking resistance anisotropy in mesoscopic LAO/STO samples with domain walls should be considered in the future study of oxide electronics and may be exploited for applications.

**Results**

In this study, characterization of LAO/STO is performed on samples grown by pulsed laser deposition (PLD). Most of samples studied had 10 unit cell (10 u.c.) LAO and were deposited at $O_2$ partial pressure of $10^{-4}$ or $10^{-5}$ Torr, although an 8 u.c. sample was also studied for comparison. Fig.1a shows a typical scanning transmission electron microscopy (STEM) image recorded on a LAO/STO sample with 10 u.c. LAO film. Although a small level of cation intermixing is likely to have occurred[31], the PLD growth resulted in well-defined interfaces with no misfit dislocation detected (Fig. 1a and Supplementary Figure S1). The samples were found to have electron density in the range of $10^{13}$-$10^{15}/cm^2$ at room temperature, consistent with literature for LAO thicknesses above the critical thickness of 4 u.c.

It has been known since the 1960's that $SrTiO_3$ exhibits a cubic to a non-polar tetragonal phase transition at critical temperature $T_c \sim 105$ K, driven by the rotation of $TiO_6$ octahedron along the elongated *c*-axis, as it is cooled from room temperature[1,2,32-34]. Using polarized light microscopy, the tetragonal crystal domains in the STO substrate of LAO/STO can be imaged and characterized[2,32]. A continuous flow microscopy cryostat with an optical window allowed us to image tetragonal domains and domain wall formation in the LAO/STO samples and correlate domain formation with *in situ* transport measurements. Figure 1b shows the disappearance of tetragonal domains over a representative area upon warming up a 5 mm × 5 mm sized 10 u.c. sample between 77 K – 100 K. In these images, striped tetragonal domains are 20 - 40 μm wide, consistent with previously published results[2,32,33]. In such domain imaging experiments, striped tetragonal domains typically start to show visible disappearance around 90K and appear to be completely disappeared around 100K (Fig.1b), close to the ~85K $R(T)$ hysteresis seen in Ref.35 and the frequently cited $T_c$ ~105 K in bulk STO. Accompanying the disappearance of tetragonal domains between 90 K – 100 K is a hysteresis effect observed in the temperature dependent resistance, $R(T)$. Cool-down and warm-up resistance curves show some deviation below 90 K (Fig. 1c). We observe that while the cool-down $R(T)$ is always smooth across the cubic-tetragonal transition of STO, $R(T)$ during warm-up often shows a kink when the tetragonal domains start to disappear as shown in Fig.1c. In our study here where the $R(T)$ hysteresis around 90K is more pronounced than Ref. 35, simultaneous resistance and polarized imaging measurements reveal that such hysteresis $R(T)$ is associated with the tetragonal domain formation/disappearance and domains are not detected above 100K in polarized light microscopy similar to Ref. 32&33. We also note that in our experiments, thermally cycling the sample across the transition temperature appears to induce very little change to the domain pattern (Supplementary Figure S2), although the thermal cycling process sometimes causes the imaging focus to be slightly off the surface and artificially causes the domain pattern to change. The consistent domain pattern upon thermal cycling suggests that instead of being random, the domain nucleation process is likely dominated by fixed crystalline defects in our samples. This 'memory' effect of domain walls originated from inherent defects in the STO was also seen in a recent paper (Ref. 34) although in Ref.34, many finer domain patterns were found to change upon thermal cycling to room *T*.

Under the cubic-to-tetragonal transition temperature, twin domains averaging about 30 μm wide run parallel or perpendicular to each other and are split by domain walls, as shown in Fig.1. Thus when electron transport in a macroscopic sample is measured, any unique carrier transport effects due to individual domains and domain boundaries (e.g. local conductivity modulation[32] or anisotropy) could be averaged out, explaining the generally small magnitude of the hysteresis seen in transport data around $T_c$ on large samples (Fig.1 and Ref.35). To clearly understand the intrinsic electrical transport due to domains and domain walls in LAO/STO device, the measurement area was reduced to the order of the domain size. Photolithography and wet etching techniques were performed on LAO/STO to reduce the spatial size of the conductive interface to 10 × 10 μm$^2$ - 40 × 40 μm$^2$ van der Pauw squares (Fig. 2c&d). Figure 2 (a) and (b) compare the temperature dependent resistance normalized over the 300K resistance value $R_0$, for a 10 u.c. sample before (a) and after (b) reducing the conductive area from 5×5 mm$^2$ to 40 × 40 μm$^2$ (unless noted, $R(T)$ curves shown were taken during the cool-down). $R_1$ and $R_2$ are the four-wire resistances measured with current flow directions at $90^0$ to each other. Before micro-patterning, $R_1$ and $R_2$ showed consistent temperature dependence throughout the whole temperature range covered (10-300K) (Fig.2a). Note that although taking $R(T)$ during the cool-down does not show a sharp feature at the cubic-tetragonal transition ~100K like the warm-up curve, presumably due to the slow kinetics of the nucleation and growth of domains, the log-log plot of temperature dependent resistance in Fig.2a does reveal a slope change around 100K as indicated by the black arrow and dashed line. This feature is similar to a study on the temperature dependent mobility of La-doped STO films in which a ~6meV transverse optical soft phonon mode related to the antiferrodistortive cubic-tetragonal phase transition is invoked to explain the transport data between 10-200K in large samples[36]. After patterning the sample to $40 \times 40$ μm$^2$, we observed directionally dependent anisotropy in four-wire resistance at temperatures lower than ~200K. The anisotropy also increased significantly as the temperature was lowered (Fig.2b). It is also interesting to note the opposite trend of $R(T)$ at $T<30K$ along the two directions in Fig.2b: while $R_1$ shows metallic behavior (black line), $R_2$ exhibits an upturn reminiscent of the Kondo-like scattering behavior[37,38]. This anisotropic behavior of $R(T)$ along two perpendicular directions of mesoscopic device alludes to the dominance of different scattering processes. Detailed transport measurements combined with polarized light imaging on various patterned samples revealed a correlation between such electrical anisotropy in mesoscopic devices and the existence of domain wall residing within the conductive area of sample.

Reflective polarized light microscopy was used to image the conducting van der Pauw patterns post-etching. We overlaid images of domain walls with the etched patterns to obtain a clear picture how domain walls intersected the square shaped van der Pauw samples (Fig 3a-c). By patterning the van der Pauw square parallel or at 45° angel to the [100], [010] sides of STO substrate, we obtained three types of mesoscopic devices: (i) with no domain walls (type I, Fig.3a), (ii) with domain walls at near 0° to the square's edge (type II, Fig.3b) and (iii) with domain walls 45°angle to the square's edge (type III, Fig.3c). In type I devices where no domain wall was residing in the device, such samples remained isotropic down to 10 K (Fig.3d). In all type II devices, we observed at least a ten times difference between $R_1$ and $R_2$ at $T$=10K (Fig. 3b,e and Fig.4), where the greater resistance ($R_1$) is parallel to the domain wall direction. In type III devices with an extra stripped domain residing inside the pattern at ~ 45° angle to the measurement current direction (Fig.3c), samples showed negligible anisotropy below 105 K (Fig. 3f), similar to the lack of anisotropy in samples with no domain walls. However, we note that Fig. 3d and f are only similar above $T \sim 180K$. Below this temperature, the resistivity of the device without any domain walls (Fig. 3d) drops far more rapidly. This difference could be an interesting signature of the structural transition at ~ 180K. It is also worth to point out that prior work by Schoofs *et al.* showed two structural transition induced hysteresis points at 85K and 180K[35]. We also clearly observed these two structural transitions in sample 10.3, whose data are displayed in Fig.3b and e. According to our *in situ* polarized light imaging, the ~90K hysteresis in $R(T)$ is correlated with the appearance/disappearance of tetragonal domains. This leaves the exact nature of the structural phase transition at ~180K an open question. However, since the polarized light microscopy did not detect any formation of anisotropic structures, we suspect this transition at ~180K to be some precursor transition to the cubic-tetragonal transition at lower temperature.

The strength of anisotropic resistance is best illustrated by plotting the relative ratio between parallel and perpendicular resistances. We redefine $R_{parallel}$ as the four-wire resistance measured parallel to a domain wall and $R_{perpendicular}$ as the one measured perpendicular or through a domain wall. Figure 4a presents the ratio between the normalized $R_{parallel}(T)$, i.e. $[R_{parallel}(T)/R_{parallel}(300K)]$, and the normalized $R_{perpendicular}(T)$, collected on several type II devices and compare with type I devices. The tabulated values and other related sample information are listed at Table S1 in supplementary section. Upon cooling, an increasing anisotropy is observed in type II devices. In Fig.4a, one sees that the anisotropy develops as $T$ decreases with a low $T$ ratio in the range 10-20 in sample 10.2, 10.3 and 8.1. On the other hand, sample 10.1 showed divergent behavior in $R_{parallel}(T)$ below 100K and the anisotropy ratio reached ~$10^5$ at 10K (Figure 4a, inset). In type I and III devices where there are no domain walls or where the domain walls are 45° to the edges of van der Pauw square, the anisotropy remained very low (Fig.4a and Fig.S3). Changing the thickness or deposition partial pressure did not have significant effect on the anisotropy of electrical behavior as shown by comparing samples 8.1 (8u.c. LAO, $10^{-5}$ Torr) and 10.5 (10 u.c. LAO, $10^{-5}$ Torr) to other 10 u.c. samples grown at $10^{-4}$ Torr (Fig.4a and Table S1).

**Discussion**

To compare the temperature scales of the increased electrical anisotropy with the structural phase transitions, we mark the positions of the ~90K and ~180K structural transitions as vertical dashed lines. Data in Fig.4a show that the electrical anisotropy increases with lowering $T$ in a smooth fashion and with no particularly sharp anomaly exhibited at ~90K or ~180K. However, when we compare the warm-up $R(T)$ of micro-patterned mesoscopic devices with the $R(T)$ during cool-down, hysteresis in $R(T)$ is still clearly present in devices with domain walls formed inside. For instance, the structural phase transitions induced hysteresis are seen in Fig3e for sample 10.3. For sample 10.1 which showed the largest anisotropy at low $T$, hysteretic behavior is also observed in the warm-up and cool-down $R(T)$ curves, as displayed in Fig.4b. The persistence of hysteretic $R(T)$ from structural phase transitions in mesoscopic devices shows that the much slower kinetics of the domain nucleation/growth during cool-down as compared to the more rapid 'melting' of domains upon warming up is still relevant even when individual domains with size ~10-40 μm are concerned.

Since polarized light imaging showed domain formation within those devices that exhibited anisotropy, a model explaining the electrical anisotropy should take into account the effects of domain walls on electron scattering and non-uniform carrier density/conductivity within the devices. Given the non-universal behavior of resistance anisotropy ratio *vs*. temperature found in different samples (Fig.4a) and the existence of anisotropy at temperatures higher than 200K (Fig.3e, Fig.4a), a full account of the electrical resistance anisotropy observed here will need to take into account other temperature dependent anisotropic scattering effects that are sample specific such as anisotropic phonon scattering in the tetragonal phase, surface scattering, inhomogeneous oxygen vacancy density (e.g. random oxygen vacancy clusters) or polarization across the domain wall[33,39]. In particular, the surface terraces from the none zero miscut angle of STO substrate may induce a continuous temperature dependent anisotropic resistance effect throughout the whole temperature range covered here[40] that interplays with the effects from structural phase transitions at ~180K and 90K. Note that among the four type II mesoscopic samples with ~90° domain wall in our study, sample 10.3 showed an anisotropy which continuously evolves with the temperature all the way up to 300K (Fig.3e and Fig.4a). Indeed, the $R_{parallel}$ and $R_{perpendicular}$ values for this sample differ by about seven times at 300K (Supplementary Figure S4), the largest among all the samples. This indicates a significant pre-existing anisotropy in this sample from factors other than the structural domain effect, rationalizing the observation of an evolving anisotropy well above the structural transition temperature. Finally, it is also possible that there are additional domain structures that are not resolved by our optical microscopy [34] that cause different magnitude and temperature dependence in the anisotropy found in various samples.

Motivated by the discovery of different conductivity inside a tetragonal domain vs. a domain wall as revealed in scanning SQUID (superconducting quantum interference device) and scanning electron microscopy experiments[32, 34], we used electromagnetic simulation software QuickField to gain some insights on the distribution of current flow and electrostatic potential in Van der Pauw square samples consisted of striped regions with different resistivity (Figure 5). In the first case, when a striped region with lower resistivity ($\rho_2$) representing a domain wall is sandwiched between two stripes with higher resistivity ($\rho_1$), the same (10nA) current through the sample induces about three times higher voltage in the configuration of current parallel to the stripe (Fig.5b) than the perpendicular configuration (Fig.5a), yielding $R_{parallel} / R_{perpendicular} \approx 3$. On the other hand, if the triped region in the middle is set to have higher resistivity, $R_{parallel} / R_{perpendicular}$ =5.7μV/34.0μV ~ 0.167 is expected (Fig.5c and d). This trend of having $R_{parallel} > R_{perpendicular}$ with a lower resistivity stripe inside the sample is in qualitative agreement with prior experiments indicating domain walls having higher conductivity[32, 34]. From the plots of current and voltage distributions in Fig.5, one sees that the counterintuitive result of four-probe $R_{parallel} > R_{perpendicular}$ despite domain wall being more conductive is caused by the effect of domain wall diverting current toward the voltage contacts where the resistivity is higher (Fig.5b). If a more resistive stripe is inserted in the sample under the parallel configuration, the current flow is mostly confined within the stripe between the current contacts and thus only a weak current reaches the region between voltage contact and a small voltage is probed (Fig.5d). This result from simulation gives a qualitative picture of why having more conductive domain walls [32,34] within the van der Pauw sample can yield a higher four-probe resistance value in the parallel configuration, a somewhat counterintuitive result at first sight. Similar discussions and calculations were made in another work published recently [41]. However, we note that the maximum $R_{parallel} / R_{perpendicular}$ we can induce by adjusting $\rho_1$ and $\rho_2$ for the simple geometry in Fig.5 is about 3, consistent with the recent finding in Ref. 41 where devices with five domains were numerically simulated. This ratio is much less than the experimental finding here of $R_{parallel} / R_{perpendicular}$ between 10 and $10^5$. Clearly, more complex models are needed to quantitatively account for the experimental data here. Interestingly, the experimentally measured devices containing a large number of domains in Ref. 41 consistently showed a larger resistance in the current perpendicular to the domain wall configuration, in contrast to the observations for a single domain wall case in this work. This difference highlights the richness of the transport effects related to structural domains in oxide interfaces when the length scale changes.

In conclusion, we have combined the polarized light microscopy imaging of structural domains and the electrical transport to study LAO/STO devices with mesoscopic size comparable to the width of individual tetragonal domains of STO substrate. Clear temperature dependent anisotropic resistance is observed in samples split by domain walls. At low temperatures ($T$=10K), the four-point resistance is significantly lower when domain walls are perpendicular to the direction of measurement current with the anisotropy strength typically between 10-20 and sometimes reaching almost $10^5$. These results point to the importance of considering structural inhomogeneity in understanding the complex electrical and magnetic properties of such system.

**Methods**

*Film Growth*
$LaAlO_3$ (LAO) films on $SrTiO_3$ (STO) (001) substrates with mistcut angle less than 0.5° (Crystal GmbH) were grown by Pulsed-Laser Deposition (PLD). Prior to the film growth, the $SrTiO_3$ (STO) substrates were etched with a chemical solution of ammonium fluoride and hydrofluoric acid at pH=6 to obtain a $TiO_2$-terminated surface and then annealed at 950°C for one hour in an oxygen-rich atmosphere. The surface morphology was verified with an Agilent 5500AFM. In the PLD chamber the base pressure of the chamber was $10^{-6}$ Torr and was increased to an $O_2$ partial pressure of $10^{-4}$ Torr or $10^{-5}$ Torr via an MKS Mass Flow Controller and Cold Cathode. The growth was performed at a temperature of about 750°C with an initial ramping rate of about 10 °C/min up to 300° C and then about 30 °C/min up to the deposition temperature. The LAO target was ablated using a 248 nm KrF excimer laser with a fluence of about 1.2 J/cm$^2$ and a repetition rate of 2 Hz. LAO films were grown at a rate of 9 pulses per layer and the growth rate was verified in situ by oscillations in Reflection High-Energy Electron Diffraction (RHEED) patterns. After deposition, films were cooled at an initial rate of about 10 C°/min and then a final rate of about 5 C°/min.

*Mesoscopic device fabrication*
Photolithography and wet etching techniques were used to reduce the sample size of LAO/STO samples (Fig. 2). Using an AutoGlow plasma system, descumming was performed in 0.6 Torr of $O_2$ at 50 watts for 30 seconds after creating a van der Pauw pattern of photo-resist on the sample surface. Samples were then further baked at 110°C for 20 minutes and allowed to cool slowly to reduce cracking. Wet etching was performed with buffered hydrofluoric acid for 5 – 7 minutes in order to etch away any LAO/STO interface not covered by the photoresist mask (Fig. 2c). After etching, the photoresist (S1805, MicroChem Inc.) was removed with acetone, revealing a clean LAO/STO van der Pauw pattern (Fig. 2d). A Zygo optical profilometer verified the etch depths were at least 6 nm. Electrical contacts were made to verify that no conductive behavior remained on etched areas. Comparing the resistance of the sample before and after the photo-lithography patterning and wet etching process shows that the device fabrication did not cause sample degradation or damage to the q2D electron system (Supplementary Fig. S5).

*Structural domain imaging*
Samples were placed in an RC102-CFM microscopy cryostat from Cryo Industries Inc., which were imaged using a Zeiss Axio Imager light microscope. Linear polarizer filters were used as a polarizer and analyzer for polarized transmitted light microscopy. The samples were imaged as the microscopy cryostat was cooled to 77 K using liquid nitrogen. The polarizer and analyzer filters were aligned accordingly to reveal tetragonal domain walls under 105 K. Multiple temperature cycles on many samples revealed that the domain walls do not rearrange due to thermal cycling (focusing on sample surface needs to be carefully checked between thermal cycles otherwise different domain patterns may appear due to slightly shifted focus). Electrical measurements were taken simultaneously in the microscopy cryostat using wire bonding and four-wire lock-in technique.

*Transport measurement*
Samples were also measured in a QuantumDesign PPMS in of a temperature range of 10 K – 300 K. Hall measurements were conducted by varying magnetic field at various temperatures. A 7 Hz sine wave was applied to the samples for electrical transport measurements. Typical current through sample was 1-100 nA. Standard four-wire lock-in techniques were used to measure the resistivity and Hall coefficients of all samples. Aluminum wire bonding was used to contact the LAO/STO interface.

**References**

1. Cowley, R. A. Lattice Dynamics and Phase Transitions of Strontium Titanate. *Phys. Rev.* **134**, A981, (1964).
2. Lytle, F. W. X-Ray Diffractometry of Low-Temperature Phase Transformations in Strontium Titanate. *J. Appl. Phys.* **35**, 2212 (1964).
3. Ohtomo, A. & Hwang, H. Y. A high-mobility electron gas at the $LaAlO_3/SrTiO_3$ heterointerface. *Nature* **427**, 423-426, (2004).
4. Nakagawa, N., Hwang, H.Y., Muller, D.A. Why some interfaces cannot be sharp. *Nature Materials* **5**, 204-209, (2006).
5. Mannhart, J., Blank, D.H.A., Hwang, H.Y., Millis, A.J., & Triscone, J.M. Two-dimensional electron gases at oxide interfaces. *MRS Bulletin*, **33**, 1027-1034 (2008).
6. Mannhart, J., Schlom, D.G., Oxide Interfaces-An Opportunity for Electronics. *Science* **327**, 1607-1611, (2010).
7. Hwang, H.Y., Iwasa, Y., Kawasaki, M., Keimer, B., Nagaosa, N. & Tokura, Y. Emergent phenomena at oxide interfaces. *Nature Materials* **11**, 103-113, (2012).
8. Cen, C., Thiel, S., Mannhart, J. & Levy, J. Oxide nanoelectronics on demand. *Science* **323**, 1026-1030 (2009).
9. Irvin, P. *et al.*, Rewritable nanoscale oxide photodetector. *Nature Photonics* **4**, 849-852 (2010).

10. Hosoda, M., Hikita, Y., Hwang, H. Y., Bell, C. Transistor operation and mobility enhancement in top-gated LaAlO$_3$/SrTiO$_3$ heterostructures. *Appl. Phys. Lett*. **103**, 103507 (2013).
11. Lu, H.L. *et al*. Reversible insulator-metal transition of LaAlO$_3$/SrTiO$_3$interface for nonvolatile memory. *Scientific Reports* **3**, 2870 (2013).
12. Caviglia, A.D., *et al*. Electric field control of the LaAlO$_3$/SrTiO$_3$ interface ground state. *Nature* **456**, 624-627 (2008).
13. Thiel, S., Hammerl, G., Schmehl, A., Schneider, C.W. & Mannhart, J. Tunable quasi-two-dimensional electron gases in oxide heterostructures. *Science* **313**, 1942-1945 (2006).
14. Reyren, N., *et al*. Superconducting interfaces between insulating oxides. *Science* **317**, 1196-1199 (2007).
15. Brinkman, A., *et al*. Magentic effects at the interface between non-magnetic oxides. *Nature Materials* **6**, 493-496 (2007).
16. Bert, J. A. *et al*. Direct imaging of the coexistence of ferromagnetism and superconductivity at the LaAlO3/SrTiO3 interface. *Nature Phys.***7**, 767-771, (2011).
17. Ariando *et al*., Electronic phase separation at the LaAlO$_3$/SrTiO$_3$interface. *Nature Comm*. **2**, 188 (2011).
18. Li, L., Richter, C., Mannhart, J. & Ashoori, R.C. Coexistence of magentic order and two-dimensional superconductivity at LaAlO$_3$/SrTiO$_3$ interfaces. *Nature Phys.***7**, 762-766 (2011).
19. Dikin, D.A., *et al*. Coexistence of Superconductivity and Ferromagnetism in Two Dimensions. *Phys. Rev. Lett.* **107**, 056802 (2011).
20. Ramirez, A.P., Oxide electronics emerge. *Science***315**, 1377-1378, (2007).
21. Chen, Y. Z. *et al*. A high-mobility two-dimensional electron gas at the spinel/perovskite interface of gamma-Al$_2$O$_3$/SrTiO$_3$. *Nat. Commun.***4**, 1371 (2013).
22. Ngo, T.Q. *et al*. Quasi-two-dimensional electron gas at the interface of gamma-Al$_2$O$_3$/SrTiO$_3$ heterostructures grown by atomic layer deposition. *J. Appl. Phys*. **118**, 115303 (2015).
23. Hu, C. *et al*., Voltage-controlled ferromagnetism and magnetoresistance in LaCoO$_3$/SrTiO$_3$ heterostructures. *J. Appl. Phys*. **114**, 183909 (2013).
24. Zhang, J. Y. , Hwang, J. , Raghavan, S. & Stemmer, S. Symmetry lowering in extreme-electron-density perovskite quantum wells, *Phys. Rev. Lett*. **110**, 256401 (2013).
25. Raghavan, S., Zhang, J.Y., Stemmer, S. Two-dimensional electron liquid at the (111) SmTiO$_3$/SrTiO$_3$ interface. *Appl. Phys. Lett*. **106**, 132104 (2015).
26. Siemons, W. *et al*. Origin of charge density at LaAlO$_3$ on SrTiO$_3$ heterointerfaces: Possibility of intrinsic doping. *Phys. Rev. Lett.* **98**, 196802(2007).
27. Herranz, G. *et al*. High mobility in LaAlO$_3$/SrTiO$_3$ heterostructures: Origin, dimensionality, and perspectives. *Phys. Rev. Lett.* **98**, 216803(2007).
28. Kormondy, K.J. *et al*. Quasi-two-dimensional electron gas at the epitaxial alumina/SrTiO$_3$ interface: Control of oxygen vacancies. *Jour. Appl. Phys*. **117**, 095303 (2015).
29. Warusawithana, M. P. *et al*. LaAlO$_3$ stoichiometry is key to electron liquid formation at LaAlO$_3$/SrTiO$_3$ interfaces. *Nat. Commun.* **4**, 2351(2013).

30. Chambers, S. A. *et al.* Instability, intermixing and electronic structure at the epitaxial LaAlO$_3$/SrTiO$_3$(001) heterojunction. *Surf. Sci. Rep.* **65**, 317-352, (2010).
31. Zaid, H. *et al.* Atomic-resolved depth profile of strain and cation intermixing around LaAlO$_3$/SrTiO$_3$ interfaces. *Scientific Reports*, **6**, 28118 (2016).
32. Kalisky, B. *et al.* Locally enhanced conductivity due to the tetragonal domain structure in LaAlO$_3$/SrTiO$_3$ heterointerfaces. *Nature Materials* **12**, 1091-1095, (2013).
33. Honig, M., Sulpizio, J.A., Drori, J., Joshua, A., Zeldov, E., & Ilani, S. Local electrostatic imaging of striped domainorder in LaAlO$_3$/SrTiO$_3$. *Nature Materials* **12**, 1112-1118, (2013).
34. Ma, H.J. H. *et al.*, Local Electrical Imaging of Tetragonal Domains and Field-Induced Ferroelectric Twin Walls in Conducting SrTiO$_3$. *Phys. Rev. Lett.* **116**, 257601 (2016).
35. Schoofs, F., Egilmez, M., Fix, T., MacManus-Driscoll, J.L. & Blamire, M.G. Impact of structural transitions on electron transport at LaAlO$_3$/SrTiO$_3$heterointerfaces. *Appl. Phys. Lett.* **100**, 081601 (2012).
36. Verma, A., Kajdos, A.P., Cain, T.A., Stemmer, S., Jena, D. Intrinsic mobility limiting mechanisms in lanthanum-doped strontium titanate. *Phys. Rev. Lett.* **112**, 216601 (2014).
37. Lee, M., Williams, J.R., Zhang, S., Frisbie, C.D., Goldhaber-Gordon, D., Electrolyte gate-controlled Kondo effect in SrTiO$_3$, *Phys. Rev. Lett.* **107**, 256601 (2011).
38. Han,K. *et al.*, Controlling Kondo-like scattering at the SrTiO3-based interfaces. *Sci. Rep.* **6**, 25455 (2016).
39. Zubko, P., Catalan, G., Buckley, A., Welche, P. R. L. & Scott, J. F. Strain-gradient-induced polarization in SrTiO$_3$ single crystals. *Phys Rev Lett* **99**, 167601(2007).
40. Brinks, P., Siemons, W., Kleibeuker, J.E., Koster, G., Rijnders, G., Huijben, M. Anisotropic electrical transport properties of a two-dimensional electron gas at SrTiO$_3$-LaAlO$_3$ interafaces. *Appl. Phys. Lett.* **98**, 242904 (2011).
41. Frenkel, Y. *et al.*, Anisotropic transport at the LaAlO$_3$/SrTiO$_3$ interface explained by microscopic imaging of channel-flow over SrTiO$_3$ domains, *ACS Appl. Mater. Interfaces*, **8**, 12514-12519 (2016).

**Acknowledgements**
This work is supported by Air Force Office of Scientific Research (AFOSR) Grant FA 9550-12-1-0441. The authors thank Richard L.J. Qiu for developing the basic device fabrication and electrical characterization techniques at the early stage of the project and Walter Lambrecht for useful discussions.

**Author Contributions**
H.Z. and M.H.B performed TEM imaging. N.J.G. performed polarized microscopy measurements and low-temperature transport measurements. R.A. and A.S. grew all samples. S.S. completed photolithography. N.J.G. performed wet etching, sample preparation, and data analysis. N.J.G. and X.P.A.G. prepared the manuscript with input from all of the coauthors. X.P.A. G. conceived the experiment and guided the work.

**Additional Information**
The authors have no competing financial interests.

**Manuscript Figures**

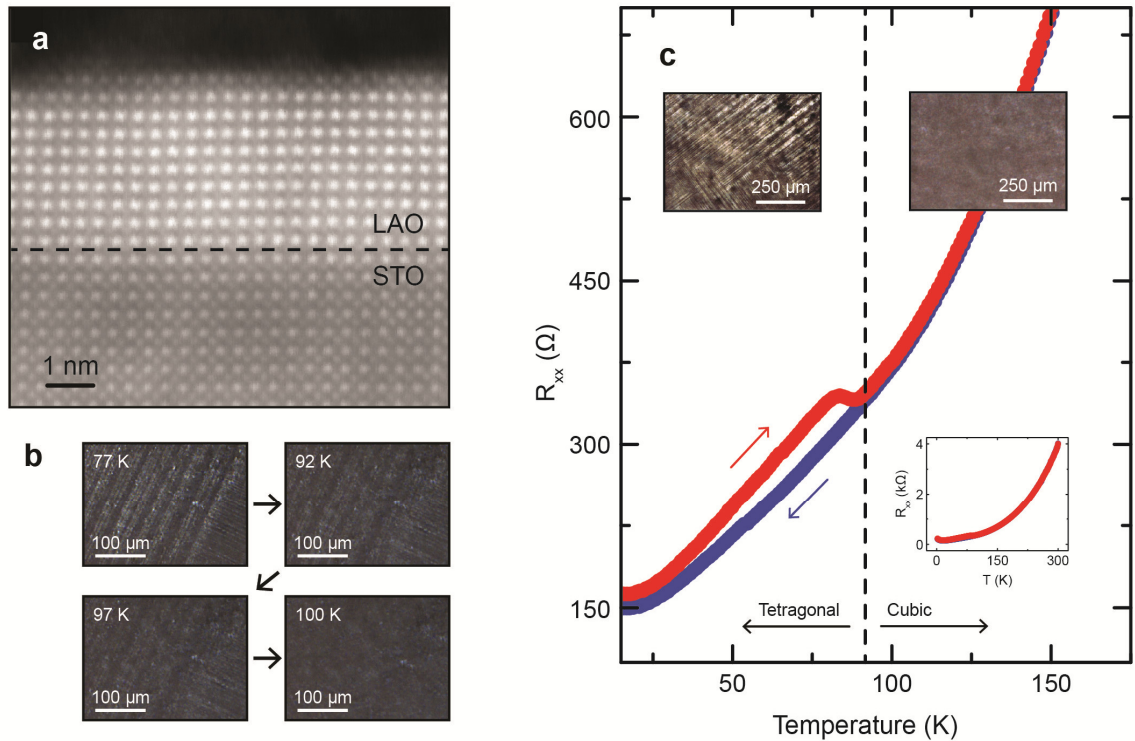

**Figure 1.** Cubic-to-tetragonal phase transition in LAO/STO and its effect on electron transport in large samples. (a) HAADF-STEM (High-Angle Annular Dark-Field Scanning Transmission Electron Microscopy) image of LAO/STO interface. Imaged sample has 10 unit cells of LAO and was grown by PLD in $10^{-4}$ Torr O$_2$ partial pressure. (b) Polarized transmission microscopy images of LAO/STO showing tetragonal domains at different temperatures. Tetragonal domain walls disappear between 97 K and 100 K. (c) Temperature dependent resistance $R(T)$ of a 5 mm×5 mm large 10 u.c. LAO/STO sample. Arrows indicate temperature sweep direction, showing hysteresis below ~90K. Insets show temperature dependent resistance from 2 K – 300 K and polarized microscopy images of an area in the sample in the tetragonal and cubic phases.

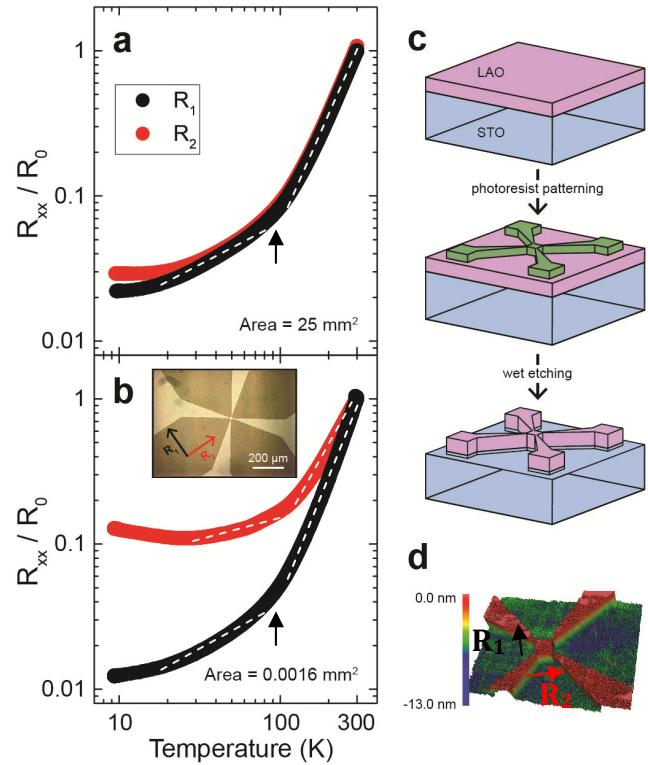

**Figure 2.** Temperature dependent resistance for a 10 u.c. LAO/STO sample before (a) and after (b) patterning. $R_1$ and $R_2$ indicate resistances taken with current and voltage contact configuration rotated by $90^0$ (see inset of (b) and (d)). Inset is a standard microscopy image of the etched van der Pauw pattern. The black arrow and white dashed lines in (a) and (b) are a guide to the eye to show the position of the cubic-tetragonal transition and the slope change in $R(T)$. The process of etching the sample is shown in **(c)**. A photoresist mask is patterned on bare LAO/STO, which is subsequently etched. The mask is removed, leaving only the conductive interface covered by the pattern. After the process is complete, the sample topography was imaged using an optical profilometer **(d)**.

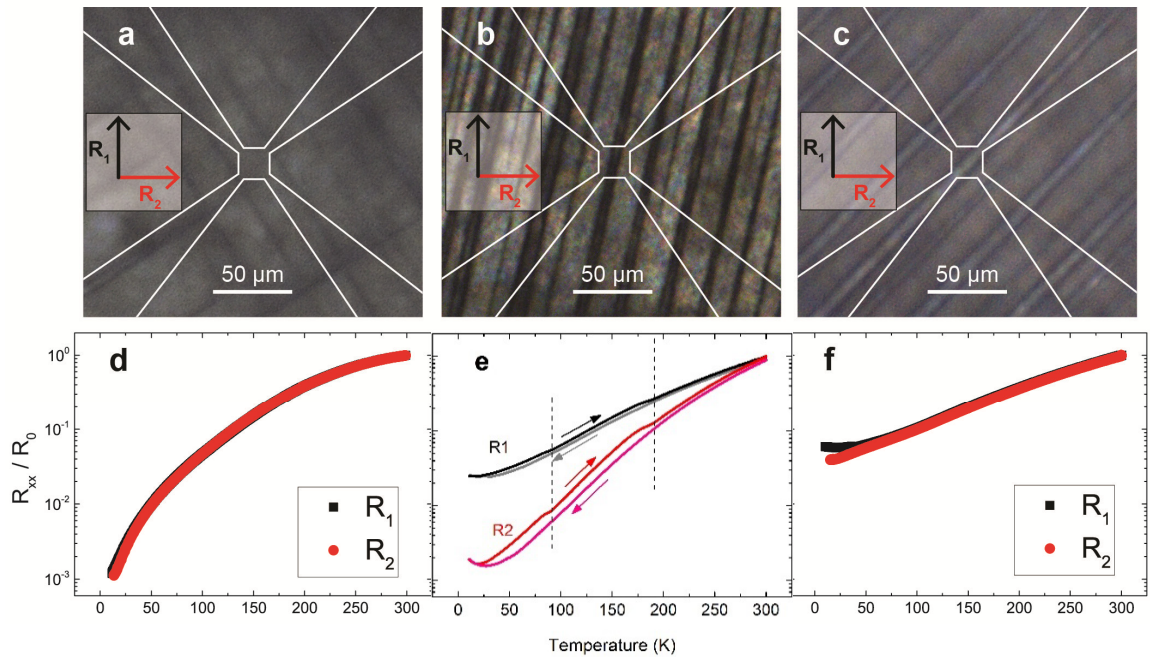

**Figure 3.** Resistance anisotropy in mesoscopic samples with domain wall in the tetragonal phase of STO. (a, b, c) Overlays of domain images and the conductive van der Pauw patterns (outlined in white). Images and data show samples at 77 K with no domain walls (a), an extra striped domain at ~80° (b), and an extra domain at ~45°(c) to the sample edge, respectively. Images (d, e, f) show the temperature dependent longitudinal resistance (normalized over the 300K value) for two measurement orientations marked in a-c. Resistance is isotropic with no domain wall (a) and 45° wall (c) and is anisotropic with 80° domain wall (b). All samples are 10u.c. thick and grown at $10^{-4}$ Torr $O_2$ pressure (samples in (a)-(c) correspond to sample no. 10.6, 10.3 and 10.4 in the Table S1 of SI).

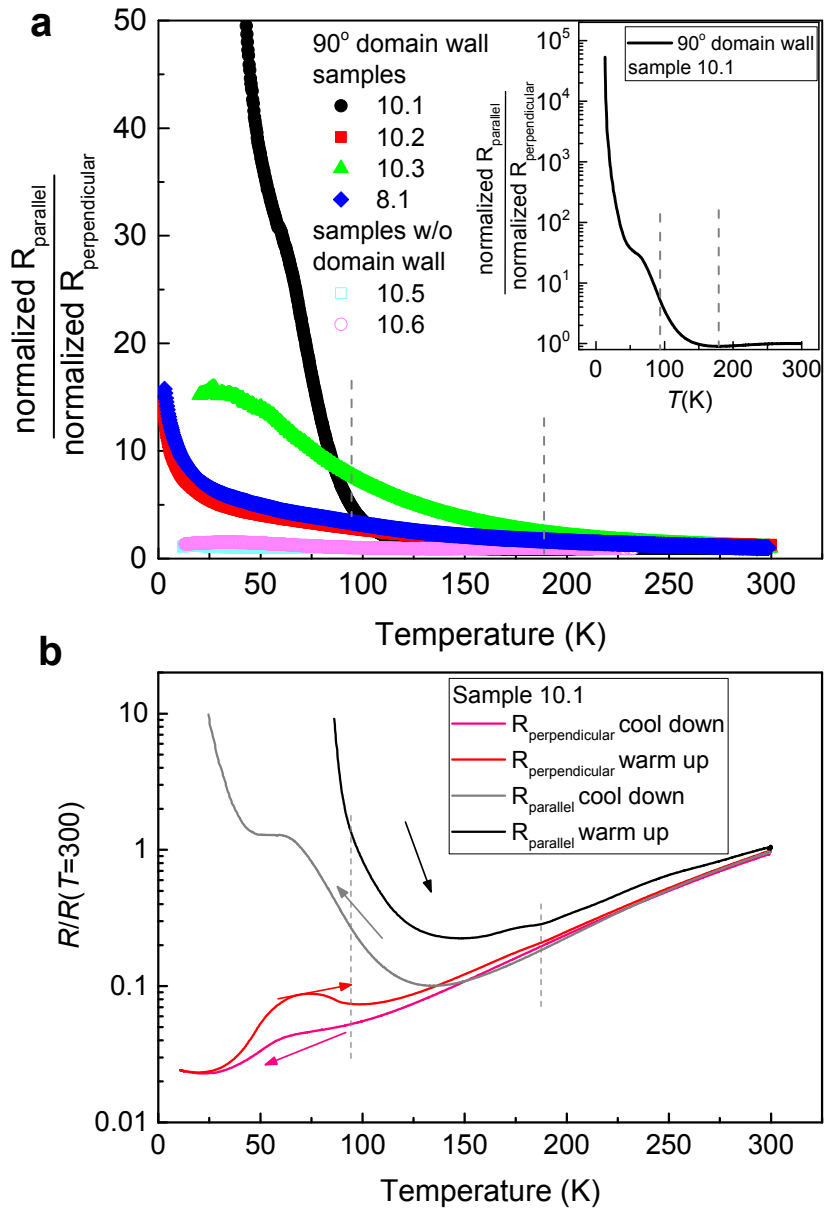

**Figure 4. (a)** Anisotropy strength in various LAO/STO samples split by one or two domain walls at ~90° vs. those without domain walls. All samples had 8–10 u.c. thick LAO and were grown at $10^{-4}$ or $10^{-5}$ Torr $O_2$ pressure. Samples with a ~90° domain wall show an anisotropy ratio between 10 and 100000 at $T=10$K. The inset shows the anisotropy ratio in log-scale for sample 10.1 which had the largest anisotropy. The two vertical dashed lines mark the temperatures for the two structural phase transitions. **(b)** The cool-down vs. warm-up $R(T)$ curves for sample 10.1 along the two measurement directions, showing the hysteresis effects.

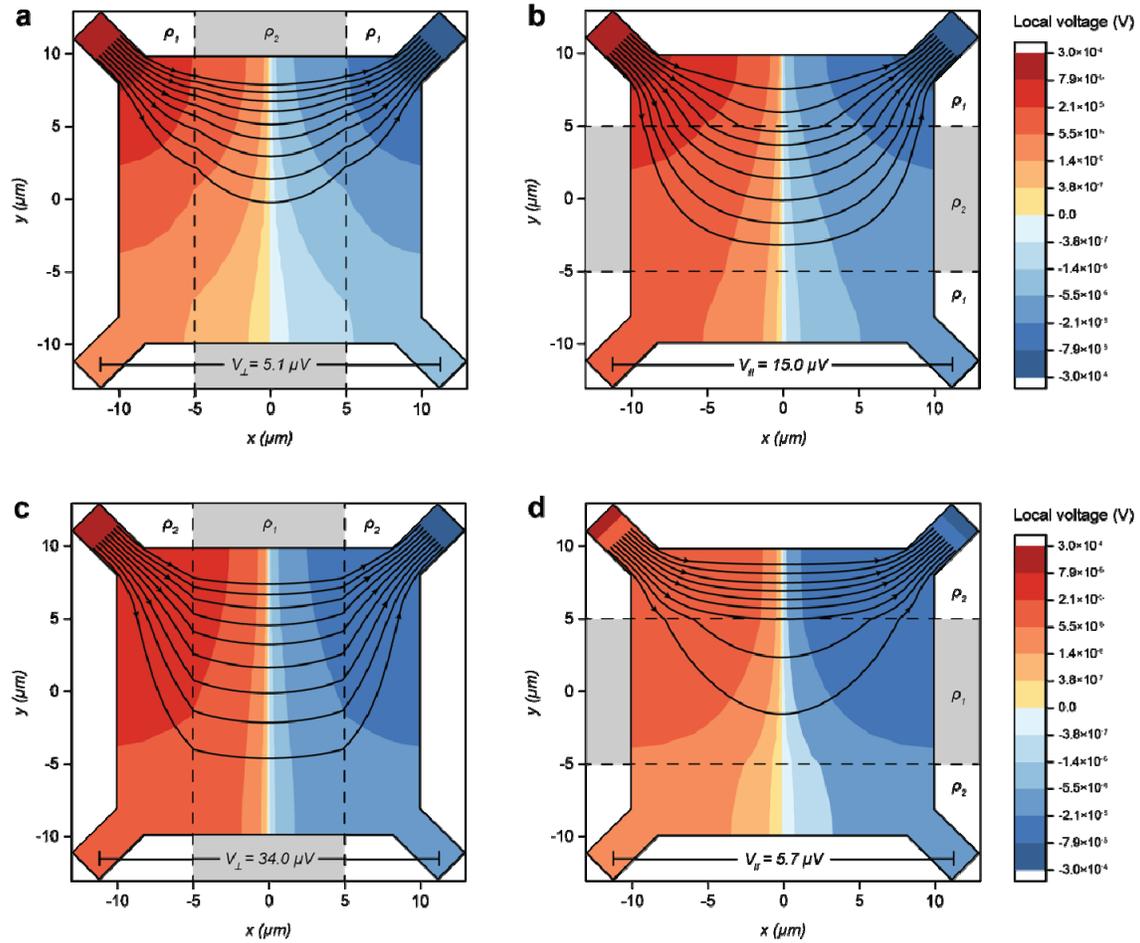

**Figure 5**. Simulated current flow and voltage distribution for inhomogeneous 20 μm × 20 μm Van der Pauw patterns. The conductive squares are split by 10 μm wide regions with different resistivity, where $\rho_1$ = 25,000 Ω and $\rho_2$ = 10,000 Ω. Total current flow is 10 nA in all simulations and is represented with streamlines. Simulated samples in (a) and (b) are split by more conductive regions oriented perpendicular and parallel to the current flow, respectively. When a conductive region splits the sample, measured $V_{parallel}$, and therefore $R_{parallel}$ (b) is greater than $V_{perpendicular}$ and $R_{perpendicular}$ (a), due to the more conducting center stripe diverting current towards voltage contacts. However, when the samples are split by a more resistive area as in (c) and (d), the sample shows smaller voltage probed by the voltage contacts in the parallel configuration (d) due to the more resistive center stripe suppressing the current flowing between voltage contacts.

# Anisotropic electrical resistance in mesoscopic LaAlO$_3$/SrTiO$_3$ devices with individual domain walls

Nicholas J. Goble[1], Richard Akrobetu[2], Hicham Zaid[3], Sukrit Sucharitakul[1], Marie-Hélène Berger[3], Alp Sehirlioglu[2], and Xuan P. A. Gao[1,*]

[1]Department of Physics, Case Western Reserve University, Cleveland, Ohio 44106, USA
[2]Department of Materials Science and Engineering, Case Western Reserve University, Cleveland, Ohio 44106, USA
[3]MINES Paris Tech, PSL Research University, MAT - Centre des matériaux, CNRS UMR 7633, BP 87 91003 Evry, France

**Supplementary Information**

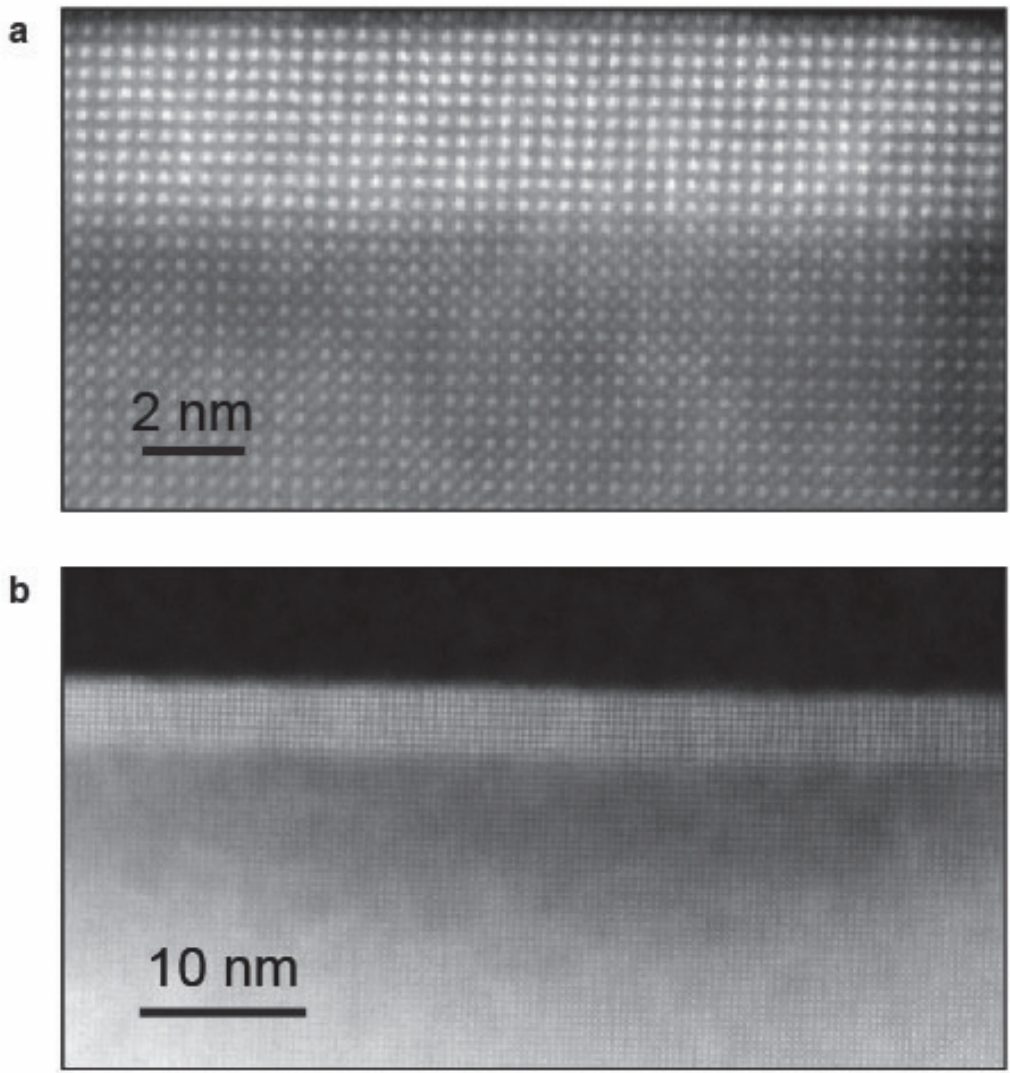

**Figure S1.** LAO/STO STEM images. Scanning transmission electron microscopy (STEM) images taken on a 10 unit cell thick LAO/STO sample grown at an oxygen partial pressure of $10^{-4}$ Torr. Brighter spots on the top of the figures are La atoms, while dimmer spots below the interface indicate Sr atoms. A clear interface forms between LAO and $TiO_2$-terminated STO.

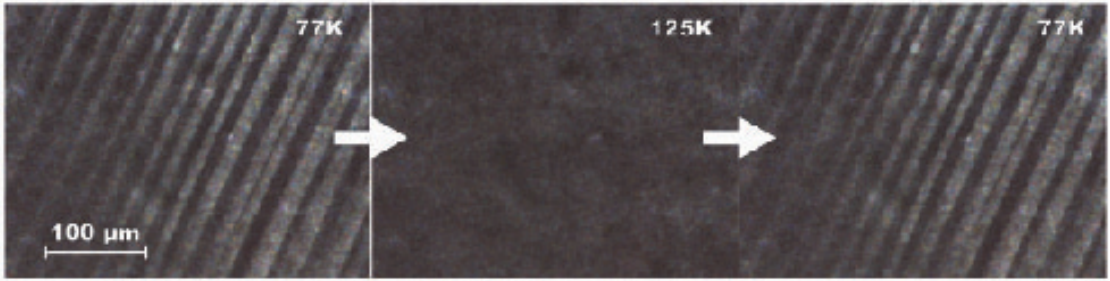

**Figure S2.** Striped tetragonal domain imaging through thermal cycling. Multiple LAO/STO samples were imaged through the structural phase transition prior to patterning and etching. Samples were heated from 77 K to 125 K and subsequently cooled back down to 77 K. We observe no change in the domain wall configuration due to thermal cycling through the cubic-to-tetragonal phase transition.

| Sample number | LAO thickness | $O_2$ partial pressure | Domain wall∠ | $R_{parallel} / R_{perp}$ (10K or 20K) | Area | Estimated carrier density (10 K) |
|---|---|---|---|---|---|---|
| 10.1 | 10 u.c. | $10^{-4}$Torr | 90° | Up to 70,000 | 40x40 µm² | $3.64 \times 10^{13}$ cm$^{-2}$ |
| 10.2 | 10 u.c. | $10^{-4}$Torr | 90° | 16 | 40x40 µm² | $2.59 \times 10^{13}$ cm$^{-2}$ |
| 10.3 | 10 u.c. | $10^{-4}$Torr | 80° | 10 | 20x20 µm² | $3.21 \times 10^{13}$ cm$^{-2}$ |
| 10.4 | 10 u.c. | $10^{-4}$Torr | 45° | 1.45 | 20x20 µm² | $3.49 \times 10^{13}$ cm$^{-2}$ |
| 10.5 | 10 u.c. | $10^{-5}$Torr | none | 1.25 | 10x10 µm² | $2.50 \times 10^{15}$ cm$^{-2}$ |
| 10.6 | 10 u.c. | $10^{-4}$Torr | none | 1.35 | 20x20 µm² | n/a |
| 8.1 | 8 u.c. | $10^{-5}$Torr | 90° | 15 | 30x30 µm² | n/a |

**Table S1.** Collection of patterned and measured samples. All seven samples studied were grown using pulsed laser deposition. A LAO thickness of 10 u.c. provided reliable interfacial conduction in all growth conditions, though a sample with 8 u.c. LAO thickness was studied as well, providing comparable results. Samples grown at a lower $O_2$ partial pressure showed substantially higher carrier densities due to the increase in oxygen vacancies in the STO crystal. Sample 10.1 was patterned with eight contacts, allowing for multiple measurement configurations across the same domain wall with different anisotropy ratios that could reach as high as 70000.

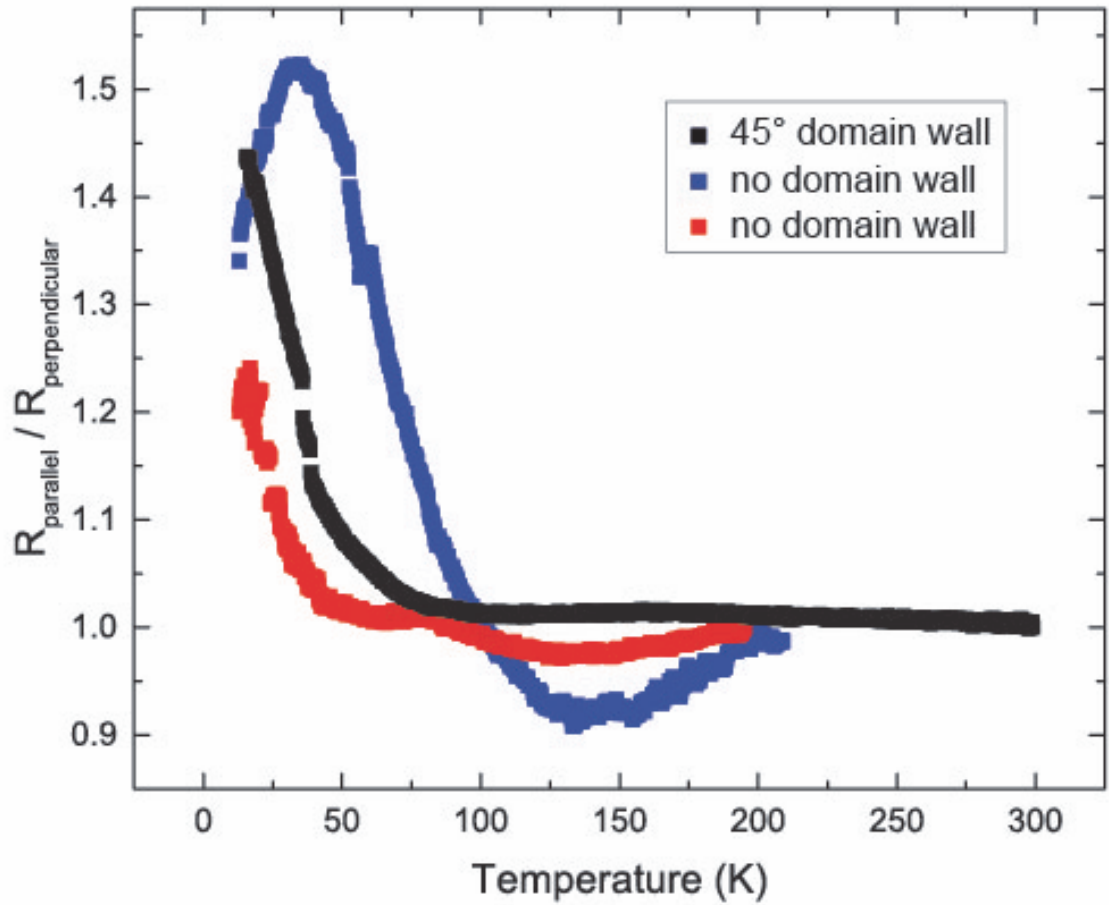

**Figure S3.** Temperature dependent anisotropy for LAO/STO samples with no domain wall or a 45° domain wall. The strength of anisotropy is minimal in samples with no domain walls. The relative difference in resistance between orthogonal measurement directions is less than 1.5 at low temperatures.

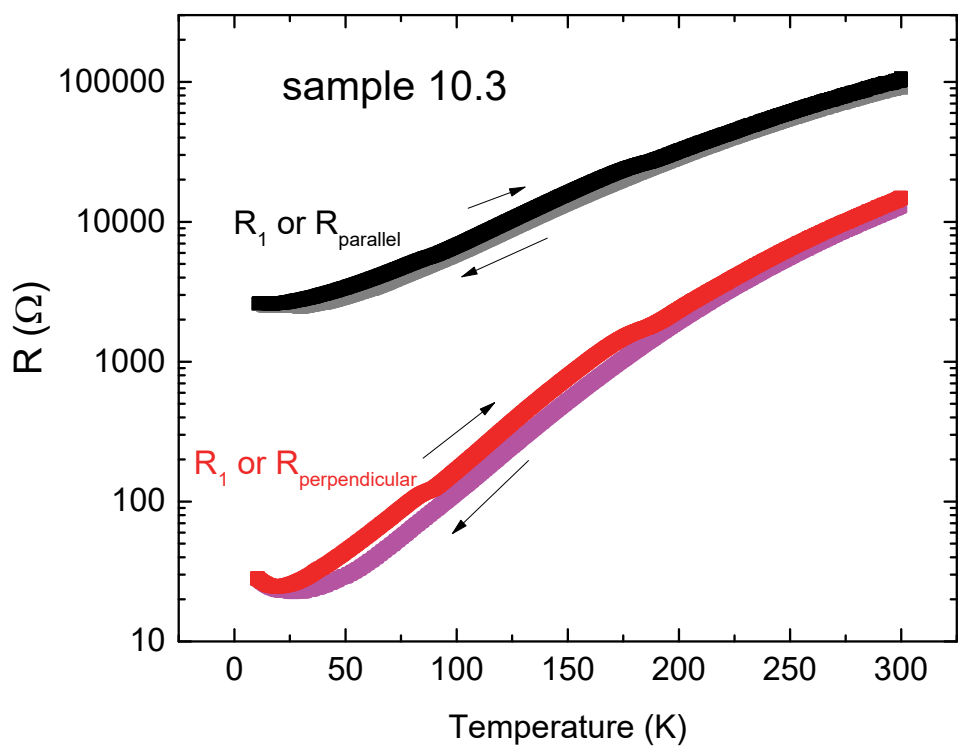

**Figure S4.** Temperature dependent resistance for LAO/STO sample 10.3 measured along two directions, showing the increasing anisotropy at low temperatures and a pre-existing anisotropy at 300K.

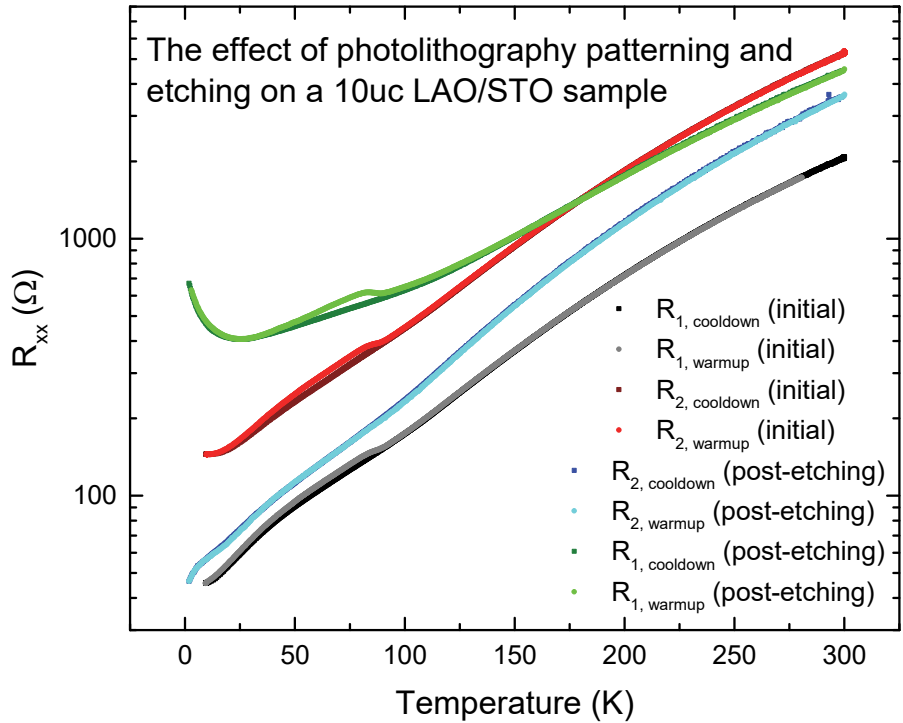

**Figure S5.** Temperature dependent resistance $R_1$ and $R_2$ of a 10uc LAO/STO sample measured along two directions at 90 degrees from each other, before (black and red dots) and after (green and blue) the photolithography patterning and wet etching. It is seen that the sample resistances before/after patterning and etching are similar at room T. At the same time, after the patterning, the difference between $R_1$ and $R_2$ at room temperature became smaller, indicating a more homogenous device after etching.

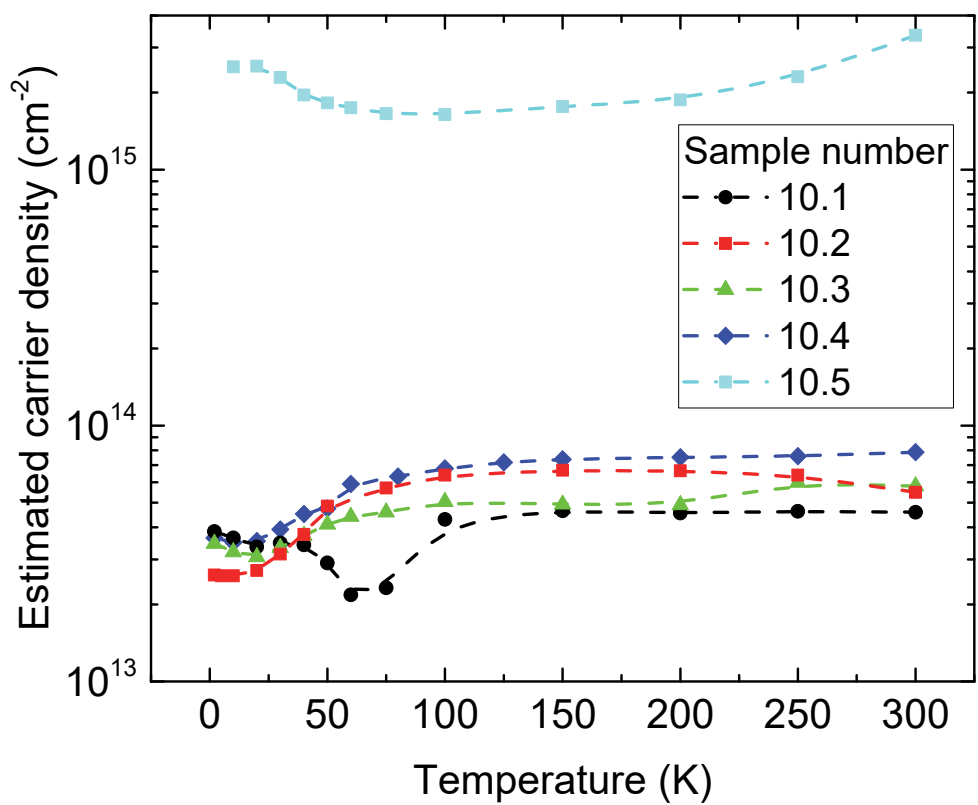

**Figure S6.** Temperature dependent carrier density for LAO/STO samples. Hall carrier density is shown throughout a temperature range of 300 K to 2 K. Values were estimated from Hall coefficients, which were collected in a ±2 T magnetic field. LAO/STO samples (#10.1-10.4) grown at a higher $O_2$ partial pressure ($10^{-4}$ Torr) have substantially less charge carriers than sample 10.5 which was grown at a lower $O_2$ partial pressure ($10^{-5}$ Torr).